\pdfoutput=1
\RequirePackage{ifpdf}
\ifpdf % We~are running pdfTeX in pdf mode
\documentclass[pdftex]{sigma}%,draft
\else
\documentclass{sigma}
\fi

\usepackage{tikz}
\usetikzlibrary{shapes,math,calc}
\usetikzlibrary{babel}

\numberwithin{equation}{section}

\newtheorem{Theorem}{Theorem}[section]
\newtheorem*{Theorem*}{Theorem}

 { \theoremstyle{definition}

 }

\begin{document}
\allowdisplaybreaks

\renewcommand{\thefootnote}{}

\newcommand{\arXivNumber}{2304.06956}

\renewcommand{\PaperNumber}{084}

\FirstPageHeading

\ShortArticleName{Non-Existence of S-Integrable Three-Point Partial Difference Equations in the Lattice Plane}

\ArticleName{Non-Existence of S-Integrable Three-Point Partial\\ Difference Equations in the Lattice Plane\footnote{This paper is a~contribution to the Special Issue on Symmetry, Invariants, and their Applications in honor of Peter J.~Olver. The~full collection is available at \href{https://www.emis.de/journals/SIGMA/Olver.html}{https://www.emis.de/journals/SIGMA/Olver.html}}}

\Author{Decio LEVI~$^{\rm a}$ and Miguel A.~RODR\'IGUEZ~$^{\rm b}$}

\AuthorNameForHeading{D.~Levi and M.A.~Rodr\'iguez}

\Address{$^{\rm a)}$~Mathematical and Physical Department, Roma Tre University,\\
\hphantom{$^{\rm a)}$}~Via della Vasca Navale, 84, I00146 Roma, Italy}

\Address{$^{\rm b)}$~Departamento de F\'{\i}sica Te\'orica, Universidad Complutense de Madrid, \\
\hphantom{$^{\rm b)}$}~Pza. de las Ciencias, 1, 28040 Madrid, Spain}
\EmailD{\href{mailto:rodrigue@ucm.es}{rodrigue@ucm.es}}

\ArticleDates{Received April 17, 2023, in final form October 23, 2023; Published online November 01, 2023}

\Abstract{Determining if an $(1+1)$-differential-difference equation is integrable or not (in the sense of possessing an infinite number of symmetries) can be reduced to the study of the dependence of the equation on the lattice points, according to Yamilov's theorem. We shall apply this result to a class of differential-difference equations obtained as partial continuous limits of 3-points difference equations in the plane and conclude that they cannot be integrable.}


\Classification{39A14; 39A36}

\begin{flushright}
\begin{minipage}{83mm}
\it This article is our contribution to the celebration\\ of Peter Olver's 70th birthday.
 We want to express our admiration and gratitude for his work and friendship.
 \end{minipage}
\end{flushright}

\renewcommand{\thefootnote}{\arabic{footnote}}
\setcounter{footnote}{0}

\section{Introduction}

Partial difference equations have always played an important role in physics and this has been noticeable the last decades. From one side, discrete systems seem to be at the base of many important laws of physics (as in quantum gravity \cite{Wi06}) and on the other side, with the increasing use of computers, discretizations are playing a growing role in physical applications to numerically solve differential equations \cite{Ha10} preserving some of the main properties of these equations, in particular their symmetries \cite{LW21}.

The number of points involved in the equation characterizes these equations, for instance, partial difference equations on four points in a plane have been studied at length \cite{AB03,GY12,LS11,LY11}, being the simplest class such that the evolution can be invertible. The continuous limit corresponds to hyperbolic partial differential equations. We can distinguish two classes of integrable equations, those which are linearizable, C-integrable equations in Calogero's terminology \cite{Ca87}, and equations whose integrability requires the solution of a scattering problem (characterized by the compatibility of two linear problems for an auxiliary wave function), S-integrable equations in Calogero's terminology.

Partial difference equations defined on three points, which appear as particular situations in triangular lattices, have not received such detailed attention (see \cite{SL13} for a thoroughly study on the linearizable case) although triangularization is a key tool in differential geometry and they allow to define many discrete models \cite{Ad00,DN03,MS01,No03,SM99}. Often, however, in the references mentioned before, the partial difference equations defined on a triangular lattice involve more than three lattice points.

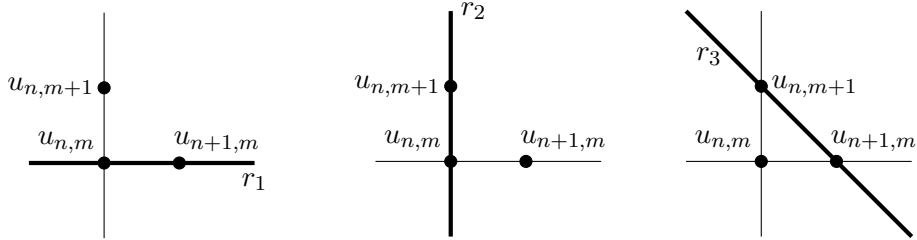
\begin{figure}[ht]\centering
\begin{tikzpicture}

\filldraw (0,0) circle (0.08);
\filldraw (1,0) circle (0.08);
\filldraw (0,1) circle (0.08);

\node[below] at (2,0) {$r_1$};

\node[above] at (1.5,0) {$u_{n+1,m}$};
\node[left] at (0,1) {$u_{n,m+1}$};
\node[above] at (-0.5,0) {$u_{n,m}$};

\draw (0,-1)--(0,2);
\draw[ultra thick] (-1,0)--(2,0);

\end{tikzpicture}
\qquad
\begin{tikzpicture}

\filldraw (0,0) circle (0.08);
\filldraw (1,0) circle (0.08);
\filldraw (0,1) circle (0.08);

\node[right] at (0,2) {$r_2$};

\node[above] at (1.5,0) {$u_{n+1,m}$};
\node[left] at (0,1) {$u_{n,m+1}$};
\node[above] at (-0.5,0) {$u_{n,m}$};

\draw[ultra thick] (0,-1)--(0,2);
\draw (-1,0)--(2,0);

\end{tikzpicture}
\qquad
\begin{tikzpicture}

\filldraw (0,0) circle (0.08);
\filldraw (1,0) circle (0.08);
\filldraw (0,1) circle (0.08);

\node[right] at (-1,1.4) {$r_3$};
\node[above] at (1.5,0) {$u_{n+1,m}$};
\node[right] at (0,1) {$u_{n,m+1}$};
\node[above] at (-0.5,0) {$u_{n,m}$};

\draw (0,-1)--(0,2);
\draw (-1,0)--(2,0);
\draw[ultra thick] (2,-1)--(-1,2);

\end{tikzpicture}
\caption{An example of points related by an equation defined on three points and the choice of initial conditions (see the main text for a discussion of these graphics). }\label{lattice}
\end{figure}

Boundary value problems can be defined for equations involving only three lattice points. In Figure~\ref{lattice}, we present just an example of the initial data and the corresponding set of points (a~more complete set of possible configurations is given in \cite{SL13}).

We consider an equation relating the values of the field $u$ at the points, $(m,n)$, $(m,n+1)$, and $(m+1,n)$ (given three points there is only one relative position, disregarding orthogonality or step sizes). Generically, if we write the equation as
\begin{equation*}
f(u_{n,m},u_{n+1,m},u_{n,m+1})=0,
\end{equation*}
we can define the value in a point as a function of the values in the other two points, for instance,
\begin{equation*}
u_{n,m+1}=g_1(u_{n,m},u_{n+1,m}),
\end{equation*}
that is, if we give the initial values along $r_1$ (see Figure~\ref{lattice}), we could compute (solving the equation) the values of the field in the upper half plane.
We can (again generically) isolate $u_{n+1,m}$:
\begin{equation*}
u_{n+1,m}=g_2(u_{n,m},u_{n,m+1}).
\end{equation*}
Then, giving the initial values along the line $r_2$ we obtain the values of the field in the right half plane. Finally, we can also obtain $u_{n,m}$ as a function of the values in the two other points:
\begin{equation*}
u_{n,m}=g_3(u_{n,m+1},u_{n+1,m}).
\end{equation*}
If we provide the initial values along the line $r_3$, the solution of the equation will be obtained in a left descending staircase. Other possible schemes can be used for this kind of lattices and equations (see \cite{Ka09} for a description of different initial-boundary value problems).

Multilinear equations can also be considered as particular cases. In fact, nonlinear equations defined on three lattice points are simpler than those on four points. The number of parameters in these equations is less than in the four points case, since the nonlinearities are at most cubic. As shown in \cite{SL13}, the classification problem for linearizable multilinear partial difference equations can be carried out up to the end and provides many examples of nonlinear three-point partial difference equations.

Let us write a generic equation on three lattice points as
\begin{equation} \label{a1}
\mathcal E_{n,m}(u_{n,m}, u_{n+1,m}, u_{n,m+1})=0.
\end{equation}
If the independent variables are written as $x_n= n h_n $ and $t_m= m h_m$, in terms of the discrete indices $n$ and $m$ and of the lattice parameters $h_n$ and $h_m$, we can get from (\ref{a1}) in the continuous limit, when $h_n,h_m \rightarrow 0$ and $n,m\rightarrow \infty$ in the usual way, equations of the form
 \begin{equation*} %\label{a2}
 u_t = f(u) u_x+ g(u) u_{xx}, \qquad u=u(x,t),
 \end{equation*}
as, for instance, the continuous Burgers equation \cite{LR83}.

The analysis of the C-integrable nonlinear partial differential equations (linearizable via a~transformation) can be found in \cite{CE87,CE87-2}. In the continuous case the class of possible transformations is richer than in the discrete one as we can have transformations of the independent and dependent variables. In the discrete case we have to restrict ourselves to just the transformations of the dependent variable unless we consider transformable lattices, i.e., partial difference schemes (see \cite{LW21} and references quoted therein).

 Contrary to the case of C-integrable equations, the situation in the case of S-integrable three-point partial difference equation is not known, up to our knowledge.

In this sense, Yamilov's theorem constitutes a useful tool in the discussion of the integrability of differential-difference equations (see, for instance, \cite{LR22} for an example of its application in a~particular case).

In fact, the classification of differential, difference or differential-difference equations with higher order symmetries, the symmetry approach to integrable equations, is a well known field with many important results (see \cite{AG79,MS87,SS90} as interesting examples of these results for differential evolution equations). In this article, we will only focus on a negative result on the existence of integrable systems of a particular type to show the role of Yamilov's theorem in this field.

The content of this article is as follows. In Section~\ref{two}, we review the main results on the linearization of partial difference equations defined on three lattice points either by point transformations or by Hopf--Cole transformations \cite{SL13}. Then in Section \ref{three}, by using a theorem introduced by Yamilov (see~\cite{LW21}), we show that a partial difference equation defined on three points cannot be S-integrable. The general result is then confirmed by considering multilinear equations where the calculations are explicit. Section \ref{four} is devoted to some concluding remarks and a proposal of future researches.

\section{C-integrable three-point partial difference equations} \label{two}

Here we review the results on linearizable three-point partial difference equations presented in~\cite{SL13}. To simplify the presentation, as we will limit ourselves to autonomous equations, we will not need to index the complex dependent variable by its lattice point but just by its relative position with respect to the reference point $n,m$, i.e. $u_{n,m}= u_{00}$. Then, equation~(\ref{a1}) can be written as
\begin{equation}
\mathcal E\left(u_{00}, u_{10}, u_{01}\right)=0.\label{Roma}
\end{equation}

In some cases, we can use an autonomous point transformation (for a nonconstant function~$f$)%
\begin{equation} \label{pp0}
\tilde u_{00}= f  (u_{00} )
\end{equation}
to linearize \eqref{Roma}, providing the linear equation
\begin{equation*}
a\tilde u_{00}+b\tilde u_{10}+c\tilde u_{01}+d=0,%\label{Janus}
\end{equation*}
 with $a$, $b$, $c$ and $d$ being complex coefficients.

 \subsection{Linearizability}

 The following theorem, see \cite{SL13}, provides a solution to this problem:
\begin{Theorem}[{\cite[Theorem 2]{SL13}}]\label{t1}
Necessary and sufficient condition for the linearizability by a~point transformation \eqref{pp0} of an equation belonging to the class \eqref{Roma} is that the equation can be written in the form
\begin{equation*}
u_{00}=\mathcal H^{-1} (-\beta\mathcal H (u_{10} )-\alpha \mathcal H (u_{01} )-\gamma ),
\end{equation*}
where $\mathcal H(x)$ is an arbitrary function of its argument, $\mathcal H^{-1}$ is its inverse function and $\alpha$, $\beta$, $\gamma$ are arbitrary integration constants. This equation is linearizable by the point transformation $\tilde u_{00}=\mathcal H (u_{00} )$ to the equation
\begin{equation*}
\tilde u_{00}+\beta\tilde u_{10}+\gamma \tilde{u}_{01}=0.
\end{equation*}
\end{Theorem}

We can also linearize \eqref{Roma} by the Cole--Hopf transformation:
\begin{equation}\label{Roma11}
\tilde u_{01}= f \left(u_{00}\right)\tilde u_{00},
\end{equation}
\noindent which transforms the linear equation
\begin{equation*}
a\tilde u_{00}+b \tilde{u}_{10}+c\tilde{u}_{01}=0,%\label{Janus1}
\end{equation*}
into (\ref{Roma}), where $a$, $b$ and $c$ are complex coefficients with $(a,b,c)\not=(0,0,0)$.
We refer to \cite{SL13} for details on the conditions over the equation to be linearizable under this transformation.

%%%%%%%%%%%%%%%%%%%%%
%%%%%%%%%%%%%%%%%%%%%
%%%%%%%%%%%%%%%%%%%%%

\subsection{Classification of linearizable multilinear equations}
We can use the above results to classify multilinear difference equations depending on three points in a two-dimensional lattice:
\begin{equation}\label{mult}
 au_{00}+ bu_{01}+ cu_{10}+ du_{00}u_{10}+ eu_{00}u_{01}+ fu_{10}u_{01}+ gu_{00}u_{10}u_{01}+ h=0,
\end{equation}
where $a,\ldots,h$ are arbitrary complex parameters. We can state the following theorem \cite{SL13}:
\begin{Theorem}[{\cite[Theorem 3]{SL13}}] Apart from the equations which are M\"obius equivalent to a~linear equation, the only equations belonging to the class \eqref{mult} which are linearizable by a~point transformation are, up to a M\"obius transformation of the dependent variable $($eventually composed with an exchange of the independent variables $n\leftrightarrow m)$, the following three
\begin{subequations}\label{Ital1-l3}
\begin{gather}
u_{00}u_{10}u_{01}-1=0,\label{Ital1}\\
u_{00}u_{01}-u_{10}=0,\label{Ital2}\\
u_{10}u_{01}-u_{00}=0.\label{Ital3}
\end{gather}
\end{subequations}
Equations \eqref{Ital1-l3} linearize by the transformation $\tilde u_{00}=\log u_{00}$ with $\log$ always standing for the principal branch of the complex logarithmic function, respectively, to the equations $(z\in \mathbb{Z})$:
\begin{gather*}
\tilde u_{00}+\tilde u_{10}+\tilde u_{01}=2\pi\mathrm{i} z,\\ %\label{Atli1}\\
\tilde u_{00}-\tilde u_{10}+\tilde u_{01}=2\pi\mathrm{i} z,\\ %\label{Atli2}\\
-\tilde u_{00}+\tilde u_{10}+\tilde u_{01}=2\pi\mathrm{i} z. %\label{Atli3}
\end{gather*}
\end{Theorem}

Let us now consider the linearization via Hopf--Cole transformations \eqref{Roma11}. As we classify up to a M\"obius transformation, we can be always set $g_{00} (u_{00} )=u_{00}$.
We can now state the following theorem:
\begin{Theorem}[{\cite[Theorem 5]{SL13}}] The class of complex autonomous multilinear discrete equations defined on three points which is linearizable to a homogeneous linear equation by a Cole--Hopf transformation $\tilde{u}_{01}=u_{00}\tilde{u}_{00}$, is given, up to a M\"obius transformation of the dependent variable $($eventually composed with an exchange of the independent variables $n\leftrightarrow m)$, by the following equation:
\begin{gather*}
\frac{1+u_{00}}{u_{00}}-\frac{1+u_{01}}{u_{10}}=0.
\end{gather*}
This equation linearizes to the equation
\begin{gather*}
\tilde{u}_{00}+\tilde{u}_{10}+\tilde{u}_{01}=0.
\end{gather*}
\end{Theorem}

\section[Yamilov's theorem and the S-integrability of three-point partial difference equations]{Yamilov's theorem and the S-integrability\\ of three-point partial difference equations}\label{three}

In \cite{LW21} (see section ``Why the shape of integrable equations on the lattice is symmetric''), the following theorem is proposed and proved.

\begin{Theorem}[Yamilov] \label{the5}
If an equation of the form
\begin{gather*}%\label{c3x}
\dot u_n = f_n = f(u_{n+N}, u_{n+N-1}, \dots, u_{n+M}),
\qquad
%\label{c3*}
N \ge M,
\qquad \frac {\partial f_n} {\partial u_{n+N}} \frac {\partial f_n}{\partial u_{n+M}} \neq 0,
\end{gather*}
where $f$ is a smooth enough function of its variables, possesses a conservation law of order $m$, such that
%\begin{equation}\label{bb14x}
$m > \min (|N|,|M|)$,
%\end{equation}
then
\begin{equation} \label{bb14y}
N = -M,
\end{equation}
and $N \ge 0$.
\end{Theorem}
This theorem states a necessary condition for a differential-difference equation to be S-integrable based on the fact that while a differential-difference equation with an infinity of symmetries is either S- or C-integrable, if it has not a sufficiently high conservation law it cannot be S-integrable. As the content of this theorem is a necessary condition, an equation which satisfies it may still be not integrable. However, a S-integrable equation has to satisfy necessarily this theorem.

Here in the following we will use this theorem to show that, while as we saw in the previous Section partial difference equations defined on three lattice points can be C-integrable, they may not be S-integrable.

To do so we carry out the partial continuous limit of \eqref{a1}. Equation~\eqref{a1} is symmetric in the exchange of $n$ and $m$ so in the whole generality we can do the continuous limit when $h_m \rightarrow 0$ and $m \rightarrow \infty$ so that $t =m h_m $ remains finite. For convenience, we will call $h_m=\varepsilon$ and $u_{n,m+1}=u_n(t+h_m)=u_n(t+\varepsilon)$. Assuming that $u_n(t)$ is an entire function of $t$, we can write
\begin{equation*} %\label{3.1}
u_n(t+\varepsilon)=u_n(t)+ \varepsilon \dot u_n(t) + \frac 1 2 \varepsilon^2 \ddot u_n(t) + \mathcal O\big(\varepsilon^3\big).
\end{equation*}
So \eqref{a1} becomes
\begin{equation} \label{3.2}
\mathcal E_{n,m}(u_{n,m},u_{n+1,m},u_{n,m+1})=\mathcal E_n(\varepsilon, u_n(t),u_{n+1}(t),u_{n}(t+\varepsilon)).
\end{equation}
Expanding the last result in \eqref{3.2} in $\varepsilon$, assuming that $\mathcal E_n(\varepsilon, u_n(t),u_{n+1}(t),u_n(t+\varepsilon))$ is an entire function, we have
\begin{gather*} %\label{3.3}
 \mathcal E_n(\varepsilon, u_n(t),u_{n+1}(t),u_n(t+\varepsilon))= \mathcal E_n^{(0)}(u_n(t),u_{n+1}(t))\!+\!
  \varepsilon  \mathcal E_n^{(1)}(u_n(t), u_{n+1}(t), \dot u_n(t))\! +\! \mathcal O\big(\varepsilon^2\big).
\end{gather*}
By a proper choice of the dependence of \eqref{a1} from $u_{m,n}$, $u_{m,n+1}$ and $u_{m+1,n}$, we can make $\mathcal E_n^{(0)}=0$ and then its semi-continuous limit becomes
\begin{equation} \label{3.4}
\mathcal E_n^{(1)}(u_n(t),u_{n+1}(t), \dot u_n(t))=0.
\end{equation}
Equation~\eqref{3.4} is not in the form of Yamilov's theorem, since there is no $u_{n-1}$ dependence in the equation, required in the theorem, and cannot be reduced to it by any lattice re-parametrization as it depends just on two points.

As an example let us consider the semi-continuous limit of \eqref{mult} when the complex parameters $a,\ldots,h$ are taken to be entire functions of $\varepsilon$, the lattice spacing in the $m$ direction which we can assume to be a constant along the lattice. We can expand the parameters of \eqref{mult} in powers of~$\varepsilon$ and we have
\[
a= a^{(0)}+ \varepsilon a^{(1)}+ \varepsilon^2 a^{(2)}+ \cdots
\]
and similar expressions for the other parameters in the equation. The request that the zero order in $\varepsilon$ of \eqref{mult} be zero implies
\begin{equation} \label{3.5}
a^{(0)}=-b^{(0)}, \qquad c^{(0)}=e^{(0)}=g^{(0)}=h^{(0)}=0, \qquad d^{(0)}=-f^{(0)},
\end{equation}
and at first order in $\varepsilon$ we get
\begin{gather}
\big(b^{(0)}+ f^{(0)}u_1\big)\dot u_0 + (a^{(1)}+b^{(1)})u_0+c^{(1)}u_1 + e^{(1)}u_0^2 + (d^{(1)}+f^{(1)})u_0 u_1\nonumber\\
\hphantom{\big(b^{(0)}+ f^{(0)}u_1\big)\dot u_0}{} + g^{(1)} u_0^2 u_1+ h^{(1)}=0.\label{3.6}
\end{gather}
Naturally \eqref{3.6} is of the form \eqref{3.4} and, by a proper choice of the parameters \eqref{3.5}, we can always make $\mathcal E_n^{(0)}=0$, as required in the general case. In the notation of Yamilov's theorem, \eqref{3.6} has $N=1$ and $M=0$ and consequently $N\ge 0$ but the condition \eqref{bb14y} is not satisfied. Thus \eqref{3.6} cannot be S-integrable as it does not satisfy the necessary S-integrability conditions implied by Theorem \ref{the5}. As~\eqref{3.6} is not S-integrable for any choice of the parameters so it will be also the multilinear partial difference equation \eqref{mult}.

\section{Conclusions}\label{four}

The application of Yamilov's theorem to the three-points partial difference equation yields
the result that no integrable differential-difference equation can be constructed by taking the limit of this class of equations.

We have just discussed the limits in the direct approach, taking the continuous limit in one of the two indices of the equation. We have also considered skew-limits (through a combination of both indices $n$, $m$), since, as it is known \cite{HJ16, LR22}, in some cases one can obtain differential-difference equations satisfying Yamilov's theorem using this approach. However, this is not possible in the three-points case.

We plan to extend the results presented here to the case of four-points partial difference equations on the lattice plane. In this case we know there are many S-integrable and C-integrable results. Since the classification of C-integrable equations in the multilinear case is not complete, it would be interesting to prove that, at least in this multilinear case, there is a privileged shape of the equation which might contain S-integrable equations.

\subsection*{Acknowledgements}
We thank the anonymous referees for corrections, useful suggestions, and constructive criticism that helped improve this article. MAR acknowledges the support of Universidad Complutense de Madrid (Spain), under grant G/6400100/3000.

{\it Professor Decio Levi passed away at the time we were working on this article. I will always miss him as my colleague and dearest friend.}

\pdfbookmark[1]{References}{ref}
\LastPageEnding


\begin{thebibliography}{99}
\footnotesize\itemsep=0pt

\bibitem{AG79}
Abellanas L., Galindo A., Conserved densities for nonlinear evolution
  equations.~{I}. {E}ven order case, \href{https://doi.org/10.1063/1.524186}{\textit{J.~Math. Phys.}} \textbf{20}
  (1979), 1239--1243.

\bibitem{Ad00}
Adler V.E., Legendre transformations on a triangular lattice, \href{https://doi.org/10.1007/BF02467062}{\textit{Funct.
  Anal. Appl.}} \textbf{34} (2000), 1--9, \href{https://arxiv.org/abs/solv-int/9808016}{arXiv:solv-int/9808016}.

\bibitem{AB03}
Adler V.E., Bobenko A.I., Suris Yu.B., Classification of integrable equations on
  quad-graphs. {T}he consistency approach, \href{https://doi.org/10.1007/s00220-002-0762-8}{\textit{Comm. Math. Phys.}}
  \textbf{233} (2003), 513--543, \href{https://arxiv.org/abs/nlin.SI/0202024}{arXiv:nlin.SI/0202024}.

\bibitem{Ca87}
Calogero F., Why are certain nonlinear {PDE}s both widely applicable and
  integrable?, in What is Integrability?, \textit{Springer Ser. Nonlinear Dynam.},
  \href{https://doi.org/10.1007/978-3-642-88703-1_1}{Springer}, Berlin, 1991, 1--62.

\bibitem{CE87}
Calogero F., Eckhaus W., Necessary conditions for integrability of nonlinear
  {PDE}s, \href{https://doi.org/10.1088/0266-5611/3/2/001}{\textit{Inverse Problems}} \textbf{3} (1987), L27--L32.

\bibitem{CE87-2}
Calogero F., Eckhaus W., Nonlinear evolution equations, rescalings, model
  {PDE}s and their integrability.~{I}, \href{https://doi.org/10.1088/0266-5611/3/2/008}{\textit{Inverse Problems}} \textbf{3}
  (1987), 229--262.

\bibitem{DN03}
Dynnikov I.A., Novikov S.P., Geometry of the triangle equation on
  two-manifolds, \href{https://doi.org/10.17323/1609-4514-2003-3-2-419-438}{\textit{Mosc. Math.~J.}} \textbf{3} (2003), 419--438,
  \href{https://arxiv.org/abs/math-ph/0208041}{arXiv:math-ph/0208041}.

\bibitem{GY12}
Garifullin R.N., Yamilov R.I., Generalized symmetry classification of discrete
  equations of a class depending on twelve parameters, \href{https://doi.org/10.1088/1751-8113/45/34/345205}{\textit{J.~Phys.~A}}
  \textbf{45} (2012), 345205, 23~pages, \href{https://arxiv.org/abs/1203.4369}{arXiv:1203.4369}.

\bibitem{Ha10}
Hansen P.C., Discrete inverse problems: {I}nsight and algorithms,
  \textit{Fundamentals of Algorithms}, Vol.~7, \href{https://doi.org/10.1137/1.9780898718836}{Society for Industrial and
  Applied Mathematics (SIAM)}, Philadelphia, PA, 2010.

\bibitem{HJ16}
Hietarinta J., Joshi N., Nijhoff F.W., Discrete systems and integrability,
  \textit{Cambridge Texts Appl. Math.}, \href{https://doi.org/10.1017/CBO9781107337411}{Cambridge University Press}, Cambridge, 2016.

\bibitem{LR83}
Levi D., Ragnisco O., Bruschi M., Continuous and discrete matrix {B}urgers'
  hierarchies, \href{https://doi.org/10.1007/BF02721683}{\textit{Nuovo Cimento~B}} \textbf{74} (1983), 33--51.

\bibitem{LR22}
Levi D., Rodr\'{\i}guez M.A., Yamilov's theorem for differential and difference
  equations, \href{https://doi.org/10.13108/2021-13-2-152}{\textit{Ufa Math.~J.}} \textbf{13} (2021), 152--159.

\bibitem{LS11}
Levi D., Scimiterna C., Linearizability of nonlinear equations on a quad-graph
  by a point, two points and generalized {H}opf--{C}ole transformations,
  \href{https://doi.org/10.3842/SIGMA.2011.079}{\textit{SIGMA}} \textbf{7} (2011), 079, 24~pages, \href{https://arxiv.org/abs/1108.3648}{arXiv:1108.3648}.

\bibitem{LW21}
Levi D., Winternitz P., Yamilov R.I., Continuous symmetries and integrability
  of discrete equations, \textit{CRM Monogr. Ser.}, Vol.~38, American
  Mathematical Society, Providence, RI, 2022.

\bibitem{LY11}
Levi D., Yamilov R.I., Generalized symmetry integrability test for discrete
  equations on the square lattice, \href{https://doi.org/10.1088/1751-8113/44/14/145207}{\textit{J.~Phys.~A}} \textbf{44} (2011),
  145207, 22~pages, \href{https://arxiv.org/abs/1011.0070}{arXiv:1011.0070}.

\bibitem{MS87}
Mikhailov A.V., Shabat A.B., Yamilov R.I., The symmetry approach to the
  classification of nonlinear equations. Complete lists of integrable systems,
  \href{https://doi.org/10.1070/RM1987v042n04ABEH001441}{\textit{Russian Math. Surveys}} \textbf{42} (1987), no.~4, 1--65.

\bibitem{MS01}
Moritz B., Schwalm W., Triangle lattice {G}reen functions for vector fields,
  \href{https://doi.org/10.1088/0305-4470/34/3/317}{\textit{J.~Phys.~A}} \textbf{34} (2001), 589--602.

\bibitem{No03}
Novikov S.P., Discrete connections on the triangulated manifolds and difference
  linear equations, \href{https://arxiv.org/abs/math-ph/0303035}{arXiv:math-ph/0303035}.

\bibitem{SM99}
Schwalm W., Moritz B., Giona M., Schwalm M., Vector difference calculus for
  physical lattice models, \href{https://doi.org/10.1103/PhysRevE.59.1217}{\textit{Phys. Rev.~E}} \textbf{59} (1999),
  1217--1233.

\bibitem{SL13}
Scimiterna C., Levi D., Three-point partial difference equations linearizable
  by local and nonlocal transformations, \href{https://doi.org/10.1088/1751-8113/46/2/025205}{\textit{J.~Phys.~A}} \textbf{46}
  (2013), 025205, 13~pages.

\bibitem{SS90}
Svinolupov S.I., Sokolov V.V., Weak nonlocalities in evolution equations,
  \href{https://doi.org/10.1007/BF01240266}{\textit{Math. Notes}} \textbf{48} (1990), 1234--1239.

\bibitem{Ka09}
van~der Kamp P.H., Initial value problems for lattice equations,
  \href{https://doi.org/10.1088/1751-8113/42/40/404019}{\textit{J.~Phys.~A}} \textbf{42} (2009), 404019, 16~pages.

\bibitem{Wi06}
Williams R.M., Discrete quantum gravity, \href{https://doi.org/10.1088/1742-6596/33/1/004}{\textit{J.~Phys. Conf. Ser.}}
  \textbf{33} (2006), 38--48.

\end{thebibliography}
\end{document}